\documentclass[12pt,preprint]{aastex}
\usepackage{psfig}

\slugcomment{KSUPT-04/4 \hspace{0.5truecm} May 2004}

\begin{document}

\title{Constraints on Scalar-Field Dark Energy from Galaxy Cluster Gas 
       Mass Fraction versus Redshift Data}

\author{Gang Chen and Bharat Ratra}

\affil{Department of Physics, Kansas State University, 116 
                 Cardwell Hall, Manhattan, KS 66506;
\mbox{chengang@phys.ksu.edu}, \mbox{ratra@phys.ksu.edu}}

\begin{abstract}

We use the Allen et al.~(2004) $Chandra$ measurements of x-ray gas mass 
fraction of 26 rich clusters to place constraints on the scalar-field 
dark energy model with inverse power law potential energy density. 
The constraints are consistent with, and typically more constraining than,
those from other cosmological tests, and mildly favor the Einstein
cosmological constant limit of the dark energy model.

\end{abstract}


\section{Introduction} 

All indications are that the energy budget of our universe has recently 
come to be dominated by some form of dark energy, resulting in an
accelerating cosmological expansion. This picture is supported by a number
of measurements, including: Type~Ia supernova redshift-magnitude 
data (see, e.g., Wang \& Mukherjee 2004; Nesseris \& Perivolaropoulos 2004; 
Riess et al.~2004; Biesiada et al.~2004; Daly \& 
Djorgovski 2004); cosmic microwave background (CMB) anisotropy data 
from the Wilkinson Microwave Anisotropy Probe (WMAP), with some input 
from other measurements (see, e.g., Bennett et al.~2003; Page et al.~2003); 
and, other measurements of CMB anisotropy, which indicate the universe 
is close to spatially flat (see, e.g., Podariu et al.~2001b; Durrer et 
al.~2003; Melchiorri \& \"Odman 2003), in combination 
with continuing strong evidence for low non-relativistic matter density 
(Chen \& Ratra 2003b and references therein). For reviews see Peebles \& 
Ratra  (2003), Padmanabhan (2003), Steinhardt (2003), Carroll (2004), and 
Sahni (2004).

There are many proposed dark energy candidates.\footnote{
For recent discussions of dark energy models and the observational
situation see Bartolo et al.~(2003), Kratochvil et al.~(2003), Kaplinghat
\& Bridle (2003), Gong (2004), Liu (2004), Gorini et al.~(2004), 
Wetterich (2004), Matarrese et al.~(2004), Bludman (2004), Feng et 
al.~(2004) and references therein.} 
The original example of dark energy is Einstein's cosmological constant, 
$\Lambda$, which has an energy density $\rho_{\Lambda}$ independent of 
time and space. The modern reincarnation of this is in the $\Lambda$CDM model
(Peebles 1984), where at low redshift non-relativistic matter, dominated
by the also hypothetical cold dark matter (CDM), is the other major 
contributor to the cosmological energy budget. More recently, dark energy
models in which the dark energy density varies only slowly with time and space
have attracted much attention. A simple example of such a candidate is a 
scalar field ($\phi$) with potential energy density $V(\phi) \propto 
\phi^{-\alpha}$, $\alpha > 0$, at low redshift $z$ (Peebles \& Ratra 
1988; Ratra \& Peebles 1988); in what follows we refer to this 
as the $\phi$CDM model. The XCDM parametrization of varying dark energy
density models approximates the dark energy by a fluid with a time-independent 
equation of state parameter $w = p/\rho$, where $p$ is the pressure.
The XCDM parameterization is accurate in the radiation and matter
dominated epochs, but much less so in the scalar field dominated epoch
when $w$ is time dependent (see, e.g., Ratra 1991).
In the last two cases, consistent with observational indications, we 
consider only a spatially flat cosmological geometry; the $\Lambda$CDM
model considered here allows space curvature to be a free parameter.

It is important to test dark energy models and constrain their parameters
using as many techniques as possible. Different tests might provide 
different constraints on the parameters of the model, and comparison of
results determined from different methods allow for consistency checks.
A number of such cosmological tests have been developed. In addition to
the Type~Ia supernova test mentioned above\footnote{ 
The proposed SNAP/JDEM space mission to measure the redshift-magnitude
relation to larger redshift should provide valuable data for constraining
dark energy models (see, http://snap.lbl.gov/ and, e.g., Podariu et 
al.~2001a; Weller \& Albrecht 2002; Rhodes et al.~2004; Virey et al.~2004).}, 
there has been discussion of the redshift-angular-size test (see, e.g., 
Zhu \& Fujimoto 2002; Chen \& Ratra 2003a; Podariu et al.~2003; Jain et 
al.~2003; Jackson 2003); the redshift-counts test (see, e.g., Loh \&
Spillar 1986; Newman \& Davis 2000; Huterer \& Turner 2001; Podariu \& 
Ratra 2001); the strong gravitational lensing test (see, e.g., Fukugita
et al.~1990; Turner 1990; Ratra \& Quillen 1992; Waga \& 
Frieman 2000; Chae 2003; Chae et al.~2004); and the redshift-expansion 
time test (see, e.g., Nolan et al.~2003; Alcaniz et al.~2003; 
Savage et al.~2004; Cepa 2004). Structure formation in 
time-variable dark energy models has also come under recent discussion 
(see, e.g., Mainini et al.~2003; Klypin et al.~2003; {\L}okas et al.~2004; 
Mota \& van de Bruck 2004), and CMB anisotropy data is providing useful 
constraints (see, e.g., Mukherjee et al.~2003a, 2003b; Caldwell \& 
Doran 2004; Wang \& Tegmark 2004; Jassal et al.~2004).

In this paper, we use the x-ray gas mass fraction of rich clusters, as a 
function of redshift, to constrain the three simple dark energy 
models mentioned above. This test was introduced by Sasaki (1996) and 
Pen (1997), and further developed by Allen et al.~(2002, hereafter A02; 
also see Allen et al.~2003; Ettori et al.~2003; Allen et al.~2004, hereafter 
A04, and references therein).\footnote{
This test builds on the zero-redshift cluster baryon mass fraction test 
discussed by White \& Frenk (1991), Fabian (1991), White (1992), and
White et al.~(1993).} 
The basic idea is as follows. Assuming that the rich clusters are large 
enough to provide a fair representation of the cosmological baryon and dark 
matter distributions (see, e.g., White 1992; White et al. 1993; Fukugita
et al.~1998), the ratio of baryonic to total mass in 
clusters---the cluster baryon mass fraction---should be the same as the 
ratio of baryonic to non-relativistic mass in the whole universe---the 
cosmological baryon mass fraction. The zero-redshift cluster baryon 
fraction test allows for a determination of the non-relativistic matter
mass density parameter $\Omega_m$ from the measured cluster baryon fraction
and an estimate of the baryonic mass density parameter $\Omega_b$. If one 
focuses on the rich clusters (and not on those that might be in the process 
of collapsing), the cluster baryon mass fraction should be independent of 
the cluster redshift. The main contributors to the cluster baryon mass 
fraction are the cluster gas mass fraction and the cluster galaxy 
(stellar) mass fraction, with the cluster gas mass fraction dominating.
The cluster gas mass fraction depends on the angular diameter distance 
(see, e.g., Sasaki 1996) so data on it as a function of redshift allows for 
another cosmological test: the correct cosmological model places the 
clusters at the right angular diameter distances to ensure that the gas
mass fractions are independent of redshift. A02 and A04 focus 
on the x-ray emitting intracluster gas, and use the x-ray gas mass fraction 
of the total cluster mass, $f_{\rm gas}$, to constrain cosmological 
parameters. Note that the optically luminous galaxy (stellar) mass in 
clusters is about $0.19 h^{0.5}$ times the x-ray emitting gas 
mass\footnote{
Here $h$ is the Hubble constant in units of 100 km s$^{-1}$ Mpc$^{-1}$.
The expression $0.19 h^{0.5}$ is from A02 and is based on Fukugita et al. 
(1998) who use a distance-independent stellar $M/L$, rather than a 
distance-dependent dynamical $M/L$, so it differs from the $0.19h^{1.5}$ 
used by White et al. (1993).},
so $\Omega_b = \Omega_m f_{\rm gas} (1 + 0.19 h^{0.5})$.

In $\S$ 2 we summarize our method of computation. Results are presented 
and discussed in $\S$ 3. We conclude in $\S$ 4.

\section{Computation}

We use the $f_{\rm gas}$ values of 26 clusters determined by A04 from 
$Chandra$ data. These are shown in Fig.~2 of A04. The redshifts of the 26
clusters range from 0.08 to 0.89.

At low $z$ the cosmological energy budget is dominated by the contributions
from non-relativistic matter, with mass density parameter $\Omega_m$, and 
dark energy, so we may ignore all other contributions. For the $\Lambda$CDM 
model and the XCDM parametrization, our analysis here is similar to that of
A02, Allen et al.~(2003), Ettori et al.~(2003), Lima et al.~(2003), 
Zhu et al.~(2004), and A04, although we differ with some of these in how we use 
priors (discussed in $\S$ 3 below). For the $\phi$CDM model we explicitly 
integrate the scalar field and other equations of motion (see, e.g., Peebles 
\& Ratra 1988) to get the needed angular diameter distance.

Following A02 and A04, we fit the $f_{\rm gas}$ data to a model described 
by
\begin{equation}
   f_{\rm gas}^{\rm mod}(z) = \frac{b\Omega_b}{(1+0.19\sqrt{h})\Omega_m}
     \left[\\
     \frac{h}{0.5}\frac{D_A^{\rm SCDM}(z)}{D_A(z,\Omega_m,p)}\right]^{3/2} ,
\end{equation}
which reflects the dependence of $f_{\rm gas}$ on the assumed angular 
diameter distance ($D_A$) to the cluster, i.e., $f_{\rm gas}\propto 
D_A^{3/2}$.\footnote{
The A04 data used in the analysis here are determined assuming an 
$\Omega_m = 1$, $\Omega_\Lambda = 0$, spatially flat, standard CDM (SCDM) 
model with $h = 0.5$, hence the $(h/0.5) D_A^{\rm SCDM}$ dependence of 
eq.~(1).}
Here $D_A$ depends on the cluster redshift $z$, $h$, and the assumed 
cosmological model, which, for each of the three cases we consider, 
depends only on two parameters, $\Omega_m$ and $p$ (where $p$ is 
$\Omega_{\Lambda}$ for the $\Lambda$CDM model, $\alpha$ for the $\phi$CDM 
model, and $\omega$ for the XCDM parametrization). 
The bias factor $b$ accounts for gas-dynamical simulation results which
indicate that the cluster baryon fraction is depressed relative
to the cosmological baryon fraction (see Allen et al.~2003; A04, and 
references therein for detailed discussions). Following A04, uncertainties 
in $\Omega_b h^2$, $h$, and $b$ are accounted for by using Gaussian priors. 
Our computations use the same Gaussian priors as A04 with $h = 0.72 
\pm 0.08$, $\Omega_b h^2 = 0.0214 \pm 0.002$, and $b = 0.824 \pm 0.089$, all 
one standard deviation errors.\footnote{
An analysis of all available measurements of the Hubble constant leads 
to a more precise estimate, $h = 0.68 \pm 0.07$ at two standard deviations
(Gott et al.~2001; Chen et al.~2003), and the value quoted
for $\Omega_b h^2$ in the main text, $\Omega_b h^2 = 0.0214 \pm 0.002$,
is more consistent with the estimate from the WMAP CMB anisotropy data
and the mean of the primordial deuterium abundance measurements, but 
significantly higher than estimates based on the primordial helium 
and lithium abundance measurements (see, e.g., Peebles \& Ratra 2003; 
Cyburt et al.~2003; Cuoco et al.~2003; Crighton et al.~2004). 
Since our analysis here is preliminary, we do not carefully investigate 
the effect of varying the chosen parameter values, although we have 
repeated the analysis using the values $h = 0.68 \pm 0.04$ and 
$\Omega_b h^2 = 0.014 \pm 0.004$ at one standard deviation (where we 
have halved the above two standard deviation error bar for $h$ and use 
the Peebles \& Ratra 2003 summary estimate for $\Omega_b h^2$; constraints
based on these numerical values are shown in the figures by using dotted 
lines and continuous lines are used to show the constraints derived using 
the favored values $h = 0.72 \pm 0.08$ and $\Omega_b h^2 = 0.0214 \pm 
0.002$) and find that the error bars on the estimated values of $\Omega_m$ 
increase by less than about 50 \%, with the most favored values of 
$\Omega_m$ becoming smaller by about a third, compared to the results 
discussed below.}

Including the above three parameters described by Gaussian priors, we need
to compute $\chi^2$ in a five dimensional parameter space, for the five
parameters $P$ = ($\Omega_m$, $p$, $h$, $\Omega_bh^2$, $b$). At a given 
point in this parameter space the $\chi^2$ difference between the model
at this point and the data is
\begin{eqnarray}
   & {} & \chi^2 (\Omega_m, p, h, \Omega_bh^2, b) = \\
      & {} & \sum_{i = 1}^{26} { \left[f_{\rm gas}^{\rm mod}(z_i, P) - 
      f_{{\rm gas},i}\right]^2 \over {\sigma_{f_{\rm gas},i}^2} } + 
      \left[{\Omega_bh^2 - 0.0214} \over {0.002} \right]^2 + 
      \left[{h - 0.72} \over {0.08}\right]^2 +
      \left[{b - 0.824} \over {0.089}\right]^2 \nonumber
\end{eqnarray}
where $f_{\rm gas}^{\rm mod}(z_i, P)$ is given in eq.~(1), and 
$f_{{\rm gas},i}$ and $\sigma_{f_{\rm gas},i}$ are the measured 
value and error from A04 for a cluster at redshift $z_i$. 

The probability distribution function (likelihood) of $\Omega_m$ and $p$ is
determined by marginalizing over the ``nuisance" parameters
\begin{equation}
   L(\Omega_m,p) = \int dh\, d(\Omega_bh^2) \, db \, e^{- \chi^2 
   (\Omega_m, p, h, \Omega_bh^2, b)/2} ,
\end{equation}
where the integral is over a large enough range of $h$, $\Omega_bh^2$, and 
$b$ to include almost all the probability. For each of the three cases
mentioned above, we compute $L(\Omega_m,p)$ on a two-dimensional grid 
spanned by $\Omega_m$ and $p$. The 1, 2, and 3 $\sigma$ confidence 
contours consist of points where the likelihood equals $e^{-2.30/2}$,
$e^{-6.17/2}$, and $e^{-11.8/2}$ of the maximum value of the likelihood, 
respectively. An alternate set of confidence contours in the 
two-dimensional ($\Omega_m$, $p$) parameter subspace may be defined by 
projecting the confidence contours (surfaces) from the five-dimensional 
($P$) parameter space.\footnote{
I.e., at each point in the two-dimensional---$\Omega_m$ and $p$---parameter 
space we adjust the values of the other three parameters---$h$, $\Omega_b
h^2$, and $b$---to minimize $\chi^2$; this defines $\chi^2 (\Omega_m, p)$ 
which is then used to derive the projected two-dimensional confidence 
contours.}
We find that the projected and marginalized contours are very similar 
and so do not show the projected contours in our plot.

\section{Results and Discussion}

In the discussion below we focus mainly on the constraints that follow 
on using $h = 0.72 \pm 0.08$ and $\Omega_b h^2 = 0.0214 \pm 0.002$, 
only noting in passing the limits that follow from using $h = 0.68 
\pm 0.04$ and $\Omega_b h^2 = 0.014 \pm 0.004$. 

The minimum value of $\chi^2$ in the full five-dimensional parameter space 
is close to 24.5 for all three models considered here. At the minimum,
$\Omega_m = 0.24$ in all three cases, and the other parameter $p$ is
$\Omega_\Lambda = 0.95$ in the $\Lambda$CDM model, $\omega = -1.2$ in the
XCDM parametrization, and $\alpha = 0$ in the $\phi$CDM model (with similar
values from the two-dimensional parameter space analyses), indicating that 
a spatially-flat $\Lambda$CDM model is somewhat favored (in all three 
cases considered this model lies within one standard deviation of the 
most likely model).

The $f_{\rm gas}$ constraints we derive on the $\Lambda$CDM model 
(Fig.~1, continuous lines) and on the XCDM parameters (Fig.~2, continuous 
lines) are quite similar to those 
in Figs.~4 and 10 of A04. As noted in A04, for 
$\Lambda$CDM these $f_{\rm gas}$ constraints are consistent with 
those from CMB anisotropy data and the brightness of distant Type Ia 
supernovae provided $\Omega_m \sim 0.3$ and $\Omega_\Lambda \sim 
0.7$. Using only the 9 clusters $f_{\rm gas}$ data of Allen et 
al.~(2003), our XCDM parametrization confidence contours are 
about twice as wide as those in Fig.~3 of Lima et al.~(2003) and 
Fig.~3 of Zhu et al.~(2004). This is because Lima et al.~ (2003) and
Zhu et al.~(2004) effectively use a Dirac-delta-like prior, akin to 
fixing $h$, $\Omega_bh^2$, and $b$ at their favored central values 
and to setting the last three terms in eq.~(2) to zero. 

Figure 3 (continuous lines) shows the  $f_{\rm gas}$ constraints on 
the $\phi$CDM model. 
These contours are tighter than those derived from the (older) Type 
Ia supernova redshift-magnitude data (Podariu \& Ratra 2000; Waga \& 
Frieman 2000), redshift-angular size data (Chen \& Ratra 2003a; Podariu 
et al.~2003), or gravitational lensing statistics (Chae et al.~2004).
The $\phi$CDM model is consistent with all of these tests when $0.15 
\lesssim \Omega_m \lesssim 0.35$ and $\alpha \lesssim 2$ (at two standard
deviations; the numerical values are modified if we instead use the 
priors of footnote 6, with a larger range of parameter space being 
acceptable). This range of parameters is also consistent with large 
scale structure and large angle CMB anisotropy data when $\Omega_bh^2$ 
is near the lower end of the allowed range (Mukherjee et al.~2003b).  

In all three figures we also show the constraints (dotted lines) that 
result from using $h = 0.68 \pm 0.04$ and $\Omega_b h^2 = 0.014 \pm 
0.004$. Comparing these contours to those discussed above (continuous
lines) gives an indication of the size of the systematic effects on
parameter determination.

\section{Conclusion}

We use recent x-ray cluster gas mass fraction data from the $Chandra$ 
observatory to constrain cosmological parameters. These constraints 
are consistent with those derived from other cosmological tests for a 
range of parameter values in each of the three cases we consider, but
mildly favor the spatially-flat $\Lambda$CDM model. The $f_{\rm gas}$ 
data is efficacious at constraining dark energy, and also provides a 
relatively tight and approximately model-independent constraint on 
$\Omega_m$, $0.15 \lesssim \Omega_m \lesssim 0.35$ at two standard 
deviations, which is in good accord with other recent estimates (Chen 
\& Ratra 2003b; Bennett et al.~2003). Future $f_{\rm gas}$ data 
should provide an even tighter constraint and is eagerly anticipated. 

\bigskip

We are indebted to S.~Allen for providing the cluster gas mass fraction 
data and helpful discussions. We also acknowledge helpful discussions 
with J.~Alcaniz, J.~Peebles, and Z.~Zhu, and support from NSF CAREER 
grant AST-9875031 and DOE EPSCoR grant DE-FG02-00ER45824. We thank the 
referee for useful advice.

\begin{figure}[p]
\psfig{file=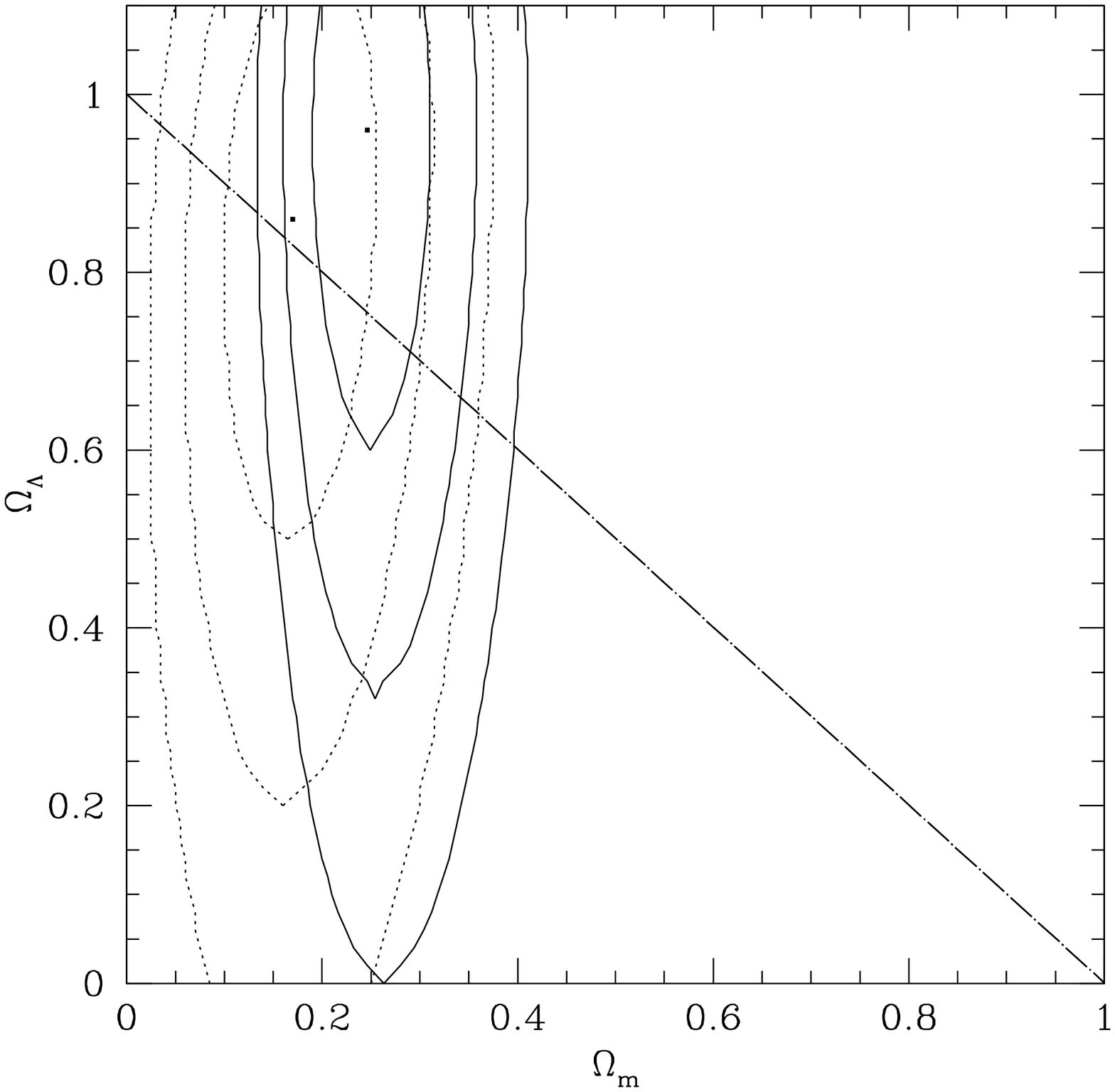,height=7.0in,width=6.7in,angle=0}
\caption{Contours of 1, 2, and 3 $\sigma$ confidence (from inside to 
outside) for the $\Lambda$CDM model. Continuous lines are computed using
$h = 0.72 \pm 0.08$ and $\Omega_b h^2 = 0.0214 \pm 0.002$ while dotted
lines are based on $h = 0.68 \pm 0.04$ and $\Omega_b h^2 = 0.014 \pm 
0.004$. The two dots denote where the likelihood is maximum. The diagonal
dot-dashed line demarcates spatially-flat models.}
\end{figure}

\begin{figure}[p]
\psfig{file=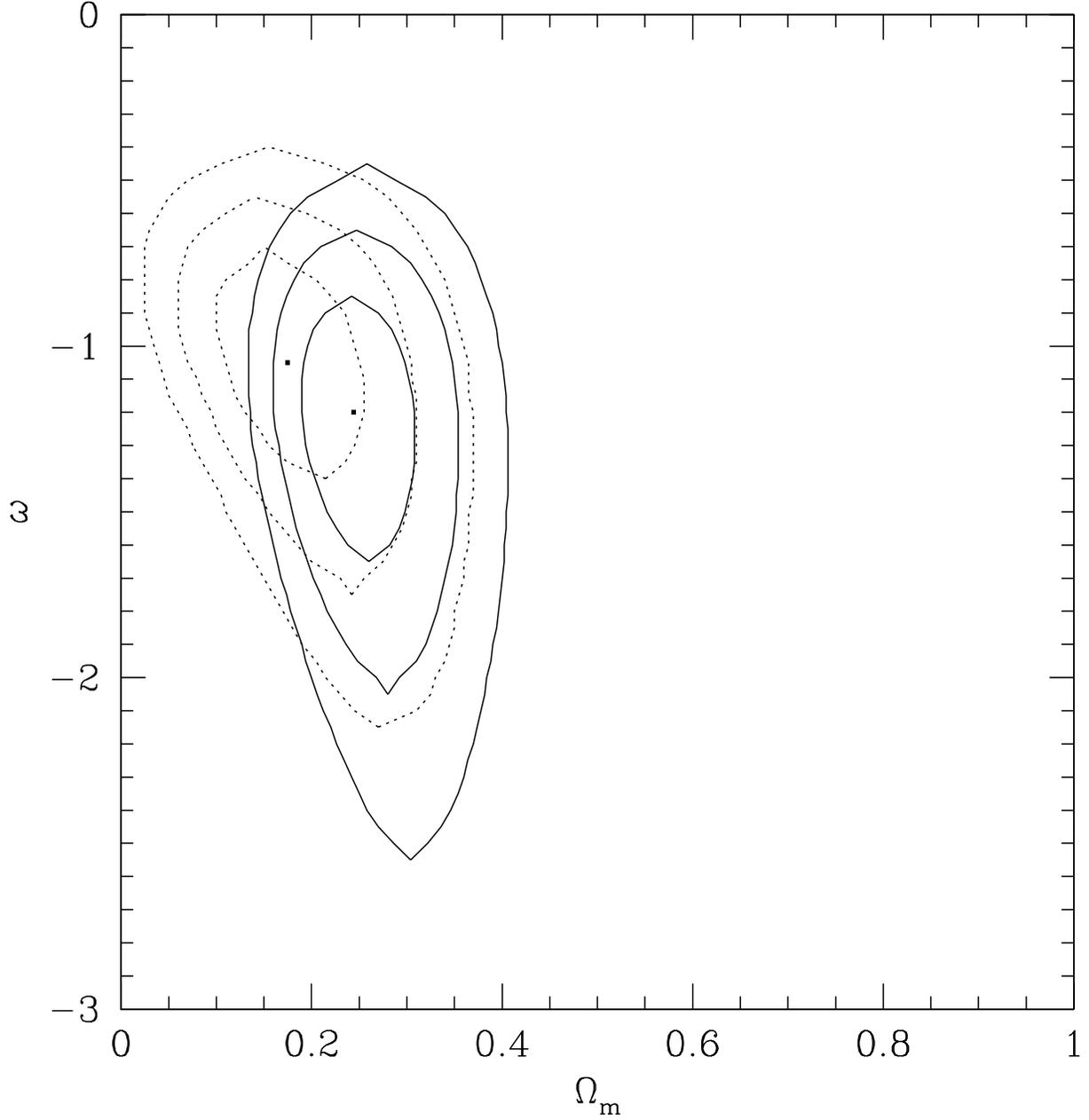,height=7.0in,width=6.7in,angle=0}
\caption{Contours of 1, 2, and 3 $\sigma$ confidence (from inside to 
outside) for the XCDM parameters. Continuous lines are computed using
$h = 0.72 \pm 0.08$ and $\Omega_b h^2 = 0.0214 \pm 0.002$ while dotted
lines are based on $h = 0.68 \pm 0.04$ and $\Omega_b h^2 = 0.014 \pm 
0.004$. The two dots denote where the likelihood is maximum.}
\end{figure}

\begin{figure}[p]
\psfig{file=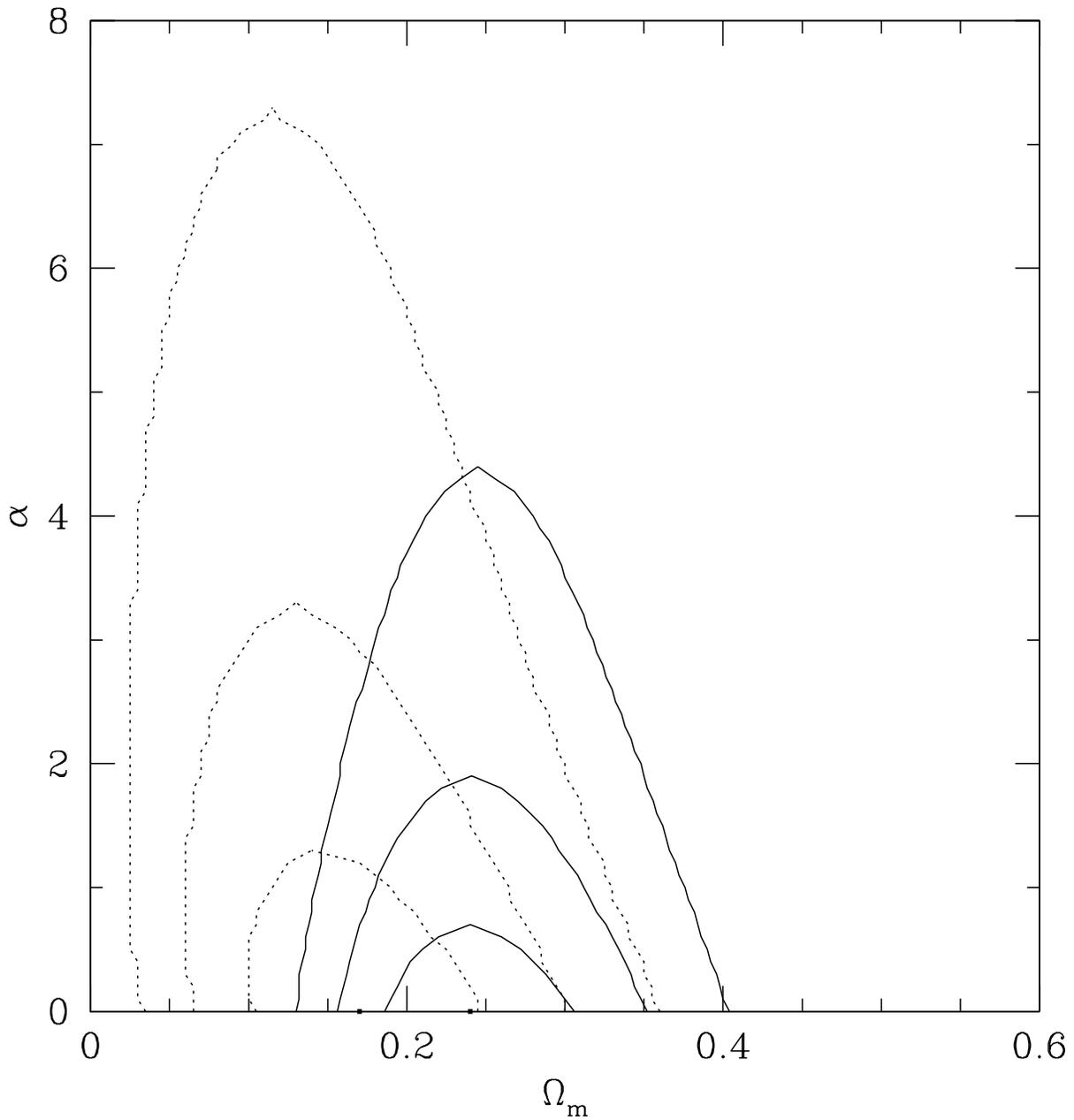,height=7.0in,width=6.7in,angle=0}
\caption{Contours of 1, 2, and 3 $\sigma$ confidence (from inside to 
outside) for the spatially-flat $\phi$CDM model with inverse power law
scalar field potential energy density $V(\phi) \propto \phi^{-\alpha}$ 
and non-relativistic CDM. Continuous lines are computed using
$h = 0.72 \pm 0.08$ and $\Omega_b h^2 = 0.0214 \pm 0.002$ while dotted
lines are based on $h = 0.68 \pm 0.04$ and $\Omega_b h^2 = 0.014 \pm 
0.004$. The two dots denote where the likelihood is maximum.}
\end{figure}

\end{document}